    \documentclass[aps,prl,twocolumn,superscriptaddress,secnumarabic]{revtex4-2}

    \usepackage{graphicx}
    \usepackage[utf8]{inputenc}
    \usepackage{float}
    \usepackage{comment}
    \usepackage[dvipsnames]{xcolor}
    \usepackage[colorlinks=true,linkcolor=OliveGreen]{hyperref}
    \hypersetup{citecolor=orange}
    \usepackage{amsmath}

\begin{document}

    \title{Multiplet-Selective Photoelectron Diffraction from an Altermagnet}

    \author{L. Plucinski}
    \email{l.plucinski@fz-juelich.de}
    \affiliation{Peter Gr{\"u}nberg Institut (PGI-6), Forschungszentrum J{\"u}lich GmbH,
    52428 J{\"u}lich, Germany}
    \affiliation{Institute for Experimental Physics II B, RWTH Aachen University, 52074 Aachen, Germany}
    
    \date{\today}

    \begin{abstract}
    
    Direct real-space probes of altermagnetic order remain scarce. Here we introduce multiplet-selective photoelectron diffraction (PED), a methodology in which different regions of a transition-metal core-level multiplet act as distinct photoemission source waves. Using multiple-scattering calculations for the metallic altermagnet candidate CrSb, we show that selected Cr $3p$ multiplet features with predominantly $Y_1^{+1}$ and $Y_1^{-1}$ character generate robust diffraction asymmetries sensitive to altermagnetic domains. We demonstrate that both circularly and linearly polarized light provide access to the effect, while suitable combinations of domains, light polarizations, and multiplet-energy windows suppress nonmagnetic diffraction backgrounds. The proposed approach can be implemented using standard momentum-resolved photoemission instrumentation and establishes core-level PED as a practical route toward domain-resolved studies of altermagnets.
    
    \end{abstract}

    \maketitle
    \section{Introduction}
    
    Altermagnets (AM) constitute a recently identified class of compensated magnetic materials in which exchange-split electronic states emerge despite vanishing net magnetization~\cite{Smejkal2022}. Unlike conventional collinear antiferromagnets (AFMs), opposite magnetic sublattices in AMs are related by rotational rather than translational symmetry, leading to momentum-dependent spin splitting and unconventional spin textures even in the absence of spin-orbit coupling.
    
    A central experimental challenge in this field is the direct detection of AM order. Recent ARPES studies on MnTe and CrSb revealed signatures broadly consistent with theoretical expectations~\cite{Krempasky2024,Hajlaoui2024,Ding2024,Li2025}, however, realistic samples contain multiple structural and magnetic domains, surface terminations, and strong matrix-element effects. In particular, circular dichroism (CD) can also arise from purely geometrical final-state effects, including the Daimon effect in photoelectron diffraction (PED)~\cite{Daimon1993}. Experimental approaches capable of separating magnetic and structural contributions therefore remain highly desirable.
    
    The present work is conceptually related to earlier magnetic photoelectron diffraction studies pioneered by Fadley and collaborators, where core-level multiplet emission from magnetic systems was used as an internal source of spin-polarized photoelectrons~\cite{Sinkovic1985,Fadley1997}. More recently, Kr\"uger demonstrated theoretically that circular dichroism in resonant photoelectron diffraction can probe staggered sublattice magnetization in AMs~\cite{Krueger2025} while recent micro-focused CD-ARPES measurements on DyMn$_6$Sn$_6$ demonstrated magnetic-domain-sensitive dichroism in the Mn $3p$ multiplet region~\cite{Plucinski2026}. Together, these developments suggest that polarization-dependent core-level PED can provide a route toward disentangling magnetic and structural contributions in AMs.
    
    In this work we investigate polarization-dependent photoelectron diffraction from CrSb surfaces. CrSb is a metallic NiAs-type AM candidate for which high-quality cleaved CrSb(001) and CrSb(120) surfaces were demonstrated experimentally~\cite{Ding2024}. In particular, the CrSb(120) surface exposes antialigned Cr sublattices embedded in inequivalent local scattering environments, making it a natural platform for multiplet-selective PED.
    
    The key ingredient of the proposed method is the transition-metal $3p$ multiplet. Different energy regions of the $3p^5 3d^n$ final-state manifold carry different mixtures of spin and orbital angular momentum~\cite{Kachel1993,Henk1999,Plucinski2026}. Consequently, selected parts of the Cr $3p$ spectrum can exhibit dominant $Y_1^{+1}$- or $Y_1^{-1}$-like character tied to the local N\'eel vector. The experiment can therefore combine several reversal operations: magnetic-domain reversal, polarization reversal, and reversal between different multiplet-energy windows. This redundancy is essential because it allows suppression of nonmagnetic diffraction backgrounds while isolating the component associated with AM order.
    
    Using polarization-dependent PED calculations~\cite{Abajo2001}, we demonstrate that both circularly and linearly polarized light generate robust sublattice-sensitive diffraction asymmetries in CrSb. Our results establish multiplet-selective core-level PED as a practical real-space probe of AM order compatible with modern micro-ARPES instrumentation.
    
    \begin{figure*}
     \centering
         \includegraphics[width=16cm]{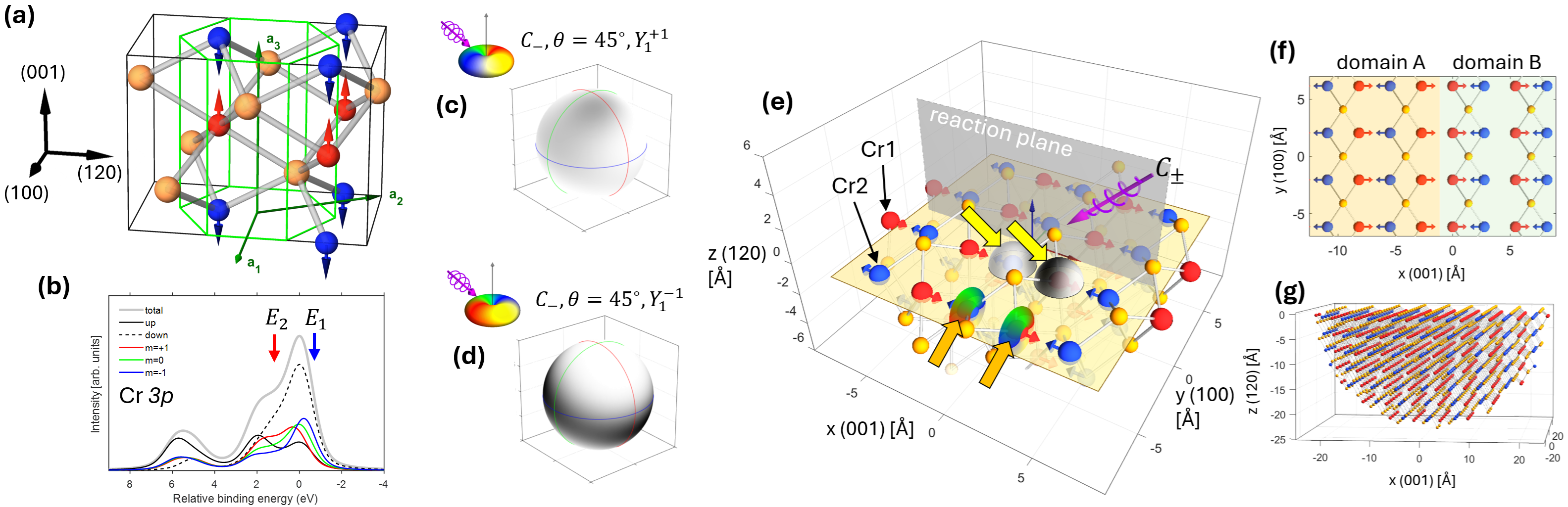}
    \caption{
    Conceptual framework of multiplet-selective photoelectron diffraction from CrSb. 
    (a) Real-space crystal and magnetic structure of CrSb. Red and blue spheres denote the two antialigned Cr sublattices (Cr1 and Cr2), while Sb atoms are shown in orange. Green and black wireframes indicate the primitive and orthorhombic unit cells, respectively.
    (b) Atomic Cr$^{3+}$ multiplet spectral function calculated using
    Quanty~\cite{Haverkort2014} for the
    $3p^6 3d^3\rightarrow3p^5 3d^3$ photoemission process, with atomic
    interaction parameters obtained from FPLO
    \cite{Koepernik1999,FPLO2202}. The thick gray curve denotes the total
    spectral function, the solid and dashed black curves denote the
    spin-resolved contributions, and the blue, green, and red curves
    denote the spin-summed $Y_1^{-1}$, $Y_1^0$, and $Y_1^{+1}$
    components, respectively. The regions $E_1$ and $E_2$ are enriched
    predominantly in $Y_1^{-1}$ and $Y_1^{+1}$ source-wave character.
    (c,d) Atomic Cr $3p$ photoionization angular distributions calculated
    using EDAC~\cite{Abajo2001} for idealized
    (c) $Y_1^{-1}$ and (d) $Y_1^{+1}$ source states. The calculations use
    $C_-$ light at $E_{\mathrm{kin}}=80$~eV, incident at an angle of
    $45^\circ$ relative to the orbital quantization axis. These panels
    illustrate the distinct dipole-emission profiles of the two pure source
    channels; the experimentally relevant $E_1$ and $E_2$ multiplet
    regions are enriched in, but are not pure realizations of, the
    corresponding $Y_1^m$ components.
    (e) CrSb(120) surface geometry and the corresponding photoelectron diffraction concept. The incident $C_\pm$ light defines the reaction plane. Multiplet-selective photoemission from Cr $3p$ states with dominant $Y_1^{-1}$ and $Y_1^{+1}$ source-wave character (indicated by orange arrows) generates different initial angular emission profiles on the Cr1 and Cr2 sublattices (yellow arrows). (f) Top view of the outermost atomic layers of the CrSb(120) surface showing domains $A$ and $B$. (g) Cluster model for CrSb(120) used in the PED calculations \cite{Abajo2001}.
    }
         \label{Fig1:Concept} 
    \end{figure*}

    \section{Multiplet-selective source waves and magnetic contrast}

    Figure~\ref{Fig1:Concept} summarizes the central concept of our work.
    CrSb exhibits three low-index surfaces, (001), (100), and (120), as
    shown in Fig.~\ref{Fig1:Concept}(a). The antialigned Cr1 and Cr2 sites
    carry opposite local moments and are embedded in inequivalent local
    environments. Removal of a Cr $3p$ electron produces a
    $3p^5 3d^3$ final-state multiplet shaped by the $3p$--$3d$ Coulomb
    and exchange interactions, spin--orbit coupling, and the local crystal
    field. We calculate this multiplet using Quanty~\cite{Haverkort2014},
    with atomic interaction parameters obtained from FPLO
    \cite{Koepernik1999,FPLO2202}, and project the energy-resolved removal
    spectrum onto the complex $Y_1^m$ basis.
    
    Figure~\ref{Fig1:Concept}(b) shows the resulting total spectrum,
    spin-resolved contributions, and spin-summed $Y_1^m$ components for
    one orientation of the local Cr moment. Within the principal multiplet
    envelope, the regions denoted $E_1$ and $E_2$ are enriched
    predominantly in $Y_1^{-1}$ and $Y_1^{+1}$ source-wave character,
    respectively. Reversal of the local moment exchanges these components,
    $Y_1^{-1}\leftrightarrow Y_1^{+1}$, while leaving the spin-summed
    $Y_1^0$ contribution unchanged. Similar magnetic sensitivity of the
    transition-metal $3p$ multiplet has been demonstrated experimentally
    for Mn $3p$~\cite{Plucinski2026}.

    In the conventional PED picture, a localized core level acts as a
    site-centered primary source whose emitted wave is subsequently
    multiply scattered by the surrounding atoms~\cite{Fadley1997}.
    Figures~\ref{Fig1:Concept}(c,d) show the calculated angular
    distributions for idealized Cr $3p$ $Y_1^{-1}$ and $Y_1^{+1}$ source
    states, using $C_-$ light incident at
    $\theta_{h\nu}=45^\circ$ relative to the orbital quantization axis
    \cite{Abajo2001}. The two channels exhibit strongly different
    intensities and angular emission profiles. Consequently, tuning the
    photoelectron energy to the $E_1$ or $E_2$ region changes the dominant
    source-wave character. Because the correspondence between
    $Y_1^{-1}$ and $Y_1^{+1}$ character and the local Cr moment reverses
    between the two magnetic sublattices, this energy selectivity provides
    a means of preferentially probing Cr1 or Cr2 within a given
    altermagnetic domain.

    Figure~\ref{Fig1:Concept}(e) illustrates the corresponding PED
    mechanism for the CrSb(120) surface. The orange arrows indicate the
    dominant $Y_1^{-1}$- and $Y_1^{+1}$-like source-wave components
    associated with the two antialigned Cr sublattices, while the yellow
    arrows depict the outgoing waves before multiple scattering.
    Because Cr1 and Cr2 are embedded in inequivalent local scattering
    environments at this surface, their emitted waves acquire different
    diffraction patterns. The resulting contrast therefore couples the
    multiplet-selective source-wave character to the non-translational
    relation between the two magnetic sublattices, which is the essential
    structural ingredient of altermagnetism. This mechanism should be
    applicable to other altermagnetic surfaces for which the two
    sublattices retain distinguishable local diffraction environments.

    \section{The case of CrSb(120) surface}
    
    As an example of the proposed formalism, we analyze the CrSb(120) surface in detail. This surface is experimentally accessible by cleaving \cite{Ding2024} and, unlike the (001) surface terminated by either Cr1 or Cr2 \cite{Sattigeri2023}, preserves compensated magnetic order.

    The CrSb(120) surface exhibits two AM domains, denoted $A$ and $B$ in
    Fig.~\ref{Fig1:Concept}(f), which differ by reversal of the local
    moments on the Cr1 and Cr2 sublattices. We first consider the
    multiplet-energy region $E_1$ in Fig.~\ref{Fig1:Concept}(b), which is
    enriched predominantly in $Y_1^{-1}$ source-wave character for the
    local-moment orientation shown. Experimentally, selecting this region
    corresponds to choosing the appropriate photoelectron kinetic energy
    within the Cr $3p$ multiplet.

    At the $E_1$ energy, the Cr1 site in domain $A$ and the Cr2 site in
    domain $B$ carry predominantly $Y_1^{-1}$ source-wave character,
    whereas the opposite sites carry predominantly $Y_1^{+1}$ character.
    For $C_-$ light at $E_{\mathrm{kin}}=80$~eV in the geometry of
    Fig.~\ref{Fig1:Concept}(e), the atomic photoionization probability is
    substantially larger for the $Y_1^{-1}$ than for the $Y_1^{+1}$
    channel, as illustrated in Figs.~\ref{Fig1:Concept}(c,d).
    Consequently, the measured PED pattern is preferentially weighted
    toward Cr1 in domain $A$ and Cr2 in domain $B$. Reversing the light
    helicity interchanges this weighting, so that $C_+$ preferentially
    enhances Cr2 in domain $A$ and Cr1 in domain $B$. This selectivity is
    not complete: both sublattices contribute, but with strongly different
    photoionization weights.

    A complementary reversal is obtained by selecting the $E_2$ region
    of Fig.~\ref{Fig1:Concept}(b), where the dominant source-wave
    character is changed from $Y_1^{-1}$ to $Y_1^{+1}$. At fixed light
    helicity, this approximately interchanges the relative Cr1 and Cr2
    photoionization weights in each domain.

    The multiple-scattering process converts this sublattice-selective
    photoionization weighting into a measurable angular contrast. For a
    given polarization and multiplet-energy region, Cr1 and Cr2 generate
    site-resolved PED patterns determined jointly by the source-wave
    character and the corresponding local scattering environment. Because
    the two environments at the CrSb(120) surface are inequivalent, these
    patterns are generally distinct, and their domain-dependent
    combinations provide the observable contrast considered below.

    To distinguish domains $A$ and $B$, PED patterns are measured at
    different lateral positions on regions having the same surface
    termination. We denote by
    $I_{\mathrm{Cr}j}^{D,C_\pm}(E_{\mathrm{kin}},\theta,\phi)$
    the site-resolved PED intensity from sublattice
    $j=1,2$ in domain $D=A,B$, where $\theta$ and $\phi$ are the polar
    and azimuthal emission angles. Restricting first to the outermost Cr1
    and Cr2 emitters, the total patterns are
    
    \begin{equation}
    \begin{aligned}
    I^{A,C_\pm} &=
    I_{\mathrm{Cr1}}^{A,C_\pm}
    +
    I_{\mathrm{Cr2}}^{A,C_\pm}, \\
    I^{B,C_\pm} &=
    I_{\mathrm{Cr1}}^{B,C_\pm}
    +
    I_{\mathrm{Cr2}}^{B,C_\pm},
    \end{aligned}
    \end{equation}
    
    Contributions from deeper Cr layers are considered in the
    Supplement~\cite{Suppl}.

    To isolate the AM contribution, we define the asymmetry

    \begin{equation}
    \begin{aligned}
    \Delta I_{\mathrm{AM}}
    = I^{A,C_-} - I^{B,C_-} - I^{A,C_+} + I^{B,C_+}, \\
    I_{\mathrm{AM}} = I^{A,C_-} + I^{B,C_-} + I^{A,C_+} + I^{B,C_+}, \\
    \Delta I_{\mathrm{AM}}^\% = 100 \cdot \Delta I_{\mathrm{AM}} / I_{\mathrm{AM}}
    \end{aligned}
         \label{Eq:AM} 
    \end{equation}
    
    \begin{figure}
     \centering
         \includegraphics[width=8.5cm]{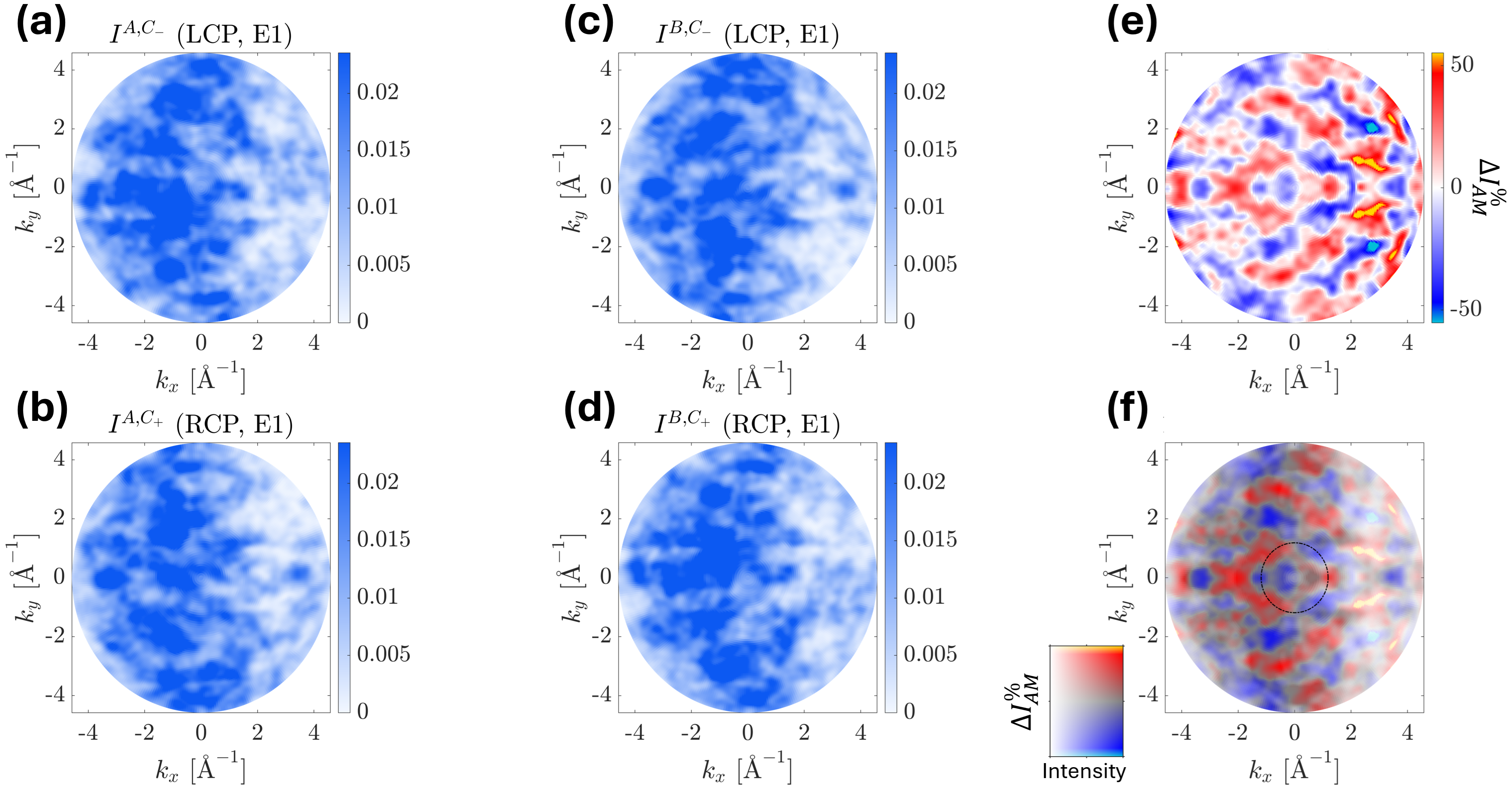}
    \caption{
    Calculated PED patterns from the CrSb(120) surface at
    $E_{\mathrm{kin}}=80$~eV for $C_\pm$ light incident at
    $\theta_{h\nu}=45^\circ$ in the geometry shown in
    Fig.~\ref{Fig1:Concept}(e). The source-wave configuration corresponds
    to the $E_1$ region defined in Fig.~\ref{Fig1:Concept}(b).
    (a--d) PED intensity maps for domains $A$ and $B$ and the two light
    helicities.
    (e) Normalized AM asymmetry $\Delta I_{\mathrm{AM}}^\%$ defined in
    Eq.~\ref{Eq:AM}.
    (f) Same asymmetry with the total intensity $I_{\mathrm{AM}}$ encoded
    through transparency. The black circle in (f) indicates the emission cone
    $\theta=\pm15^\circ$.
    }
         \label{Fig2:LCP_RCP} 
    \end{figure}

    Within the idealized source-channel model used here,
    $\Delta I_{\mathrm{AM}}$ vanishes for a conventional compensated AFM
    in which the two magnetic sublattices are related by translation. It
    also vanishes when the selected spectral region contains equal
    $Y_1^{-1}$ and $Y_1^{+1}$ source-wave weights and is therefore
    insensitive to reversal of the local moment. In the EDAC calculations
    below, the $E_1$ region is represented by its dominant
    $Y_1^{-1}$ and $Y_1^{+1}$ channels on the two oppositely magnetized
    sublattices. 
    
    To estimate the magnitude of the effect, we performed PED
    calculations using EDAC~\cite{Abajo2001,Suppl} for the cluster shown
    in Fig.~\ref{Fig1:Concept}(g). Figure~\ref{Fig2:LCP_RCP} presents the
    resulting maps for domains $A$ and $B$, the two light helicities, and
    the source-channel configuration representing the $E_1$ region.
    Although the individual patterns in
    Figs.~\ref{Fig2:LCP_RCP}(a--d) appear broadly similar, they satisfy
    the mirror relations
    $\mathcal{M}_x I^{A,C_-}=I^{B,C_+}$ and
    $\mathcal{M}_x I^{A,C_+}=I^{B,C_-}$.
    Consequently, both $I^{A,C_-}+I^{B,C_+}$ and
    $I^{A,C_+}+I^{B,C_-}$ are even under $\mathcal{M}_x$, and so is
    $\Delta I_{\mathrm{AM}}^\%$. The calculated asymmetry reaches
    approximately $30\%$ over broad angular regions, with the strongest
    contrast overlapping regions of substantial PED intensity, as shown
    in Fig.~\ref{Fig2:LCP_RCP}(f).

\begin{figure}
     \centering
         \includegraphics[width=\columnwidth]{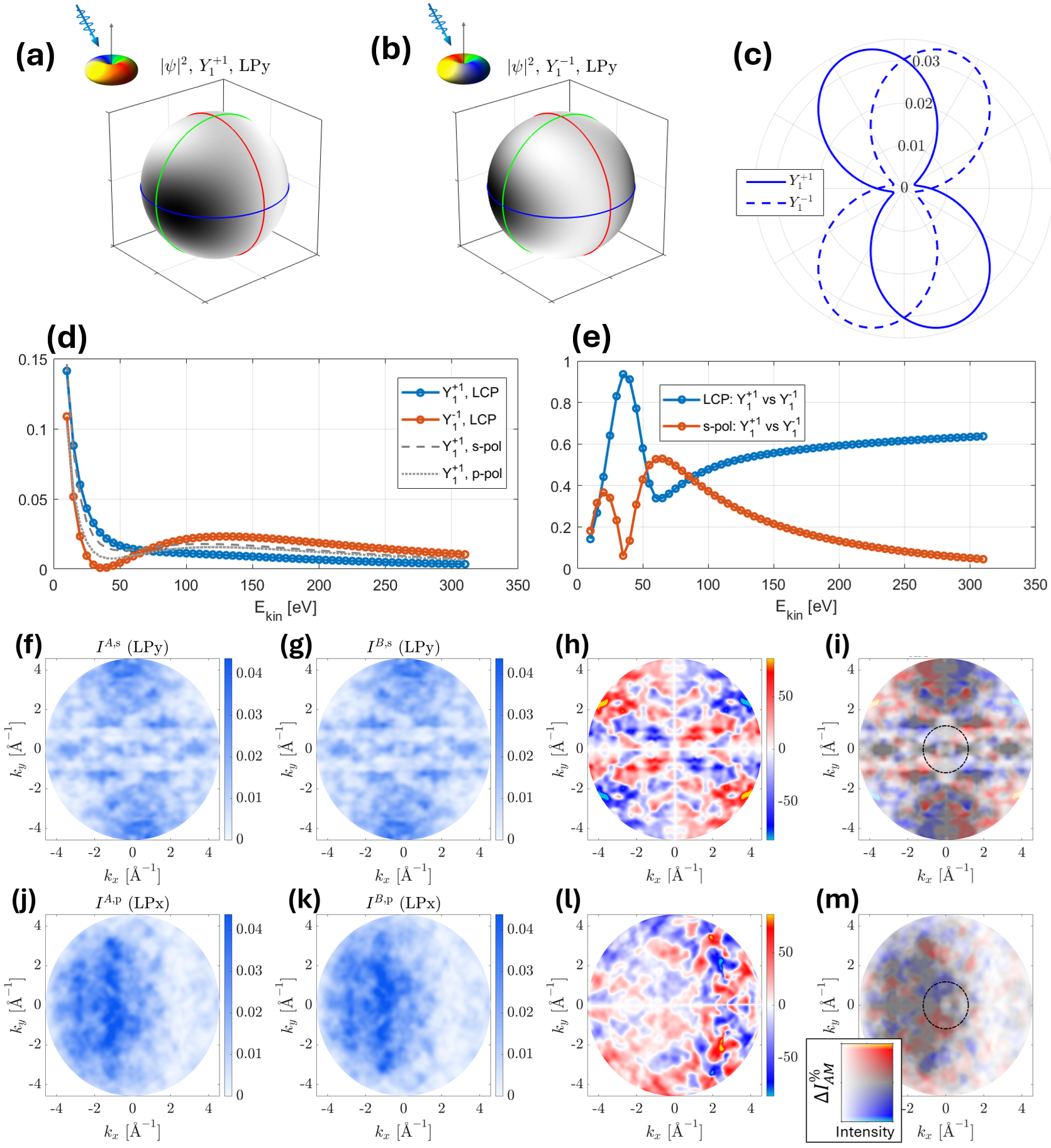}
    \caption{
    (a,b) Atomic photoionization angular distributions calculated using
    EDAC~\cite{Abajo2001} for idealized Cr $3p$ $Y_1^{-1}$ and
    $Y_1^{+1}$ source states with linearly polarized light at
    $E_{\mathrm{kin}}=80$~eV. The polarization vector lies perpendicular
    to the orbital quantization axis.
    (c) Profiles along the horizontal equators indicated in (a,b).
    (d) Kinetic-energy dependence of the integrated photoionization
    intensities for the indicated source states and light polarizations.
    (e) Kinetic-energy dependence of the integrated angular-profile
    differences for $C_-$ and $s$-polarized light.
    (f--m) Calculated PED maps from CrSb(120) at
    $E_{\mathrm{kin}}=80$~eV. Panels (f--i) correspond to $s$-polarized
    light and panels (j--m) to $p$-polarized light in the geometry of
    Fig.~\ref{Fig1:Concept}(e).
    (f,g) PED intensity maps for domains $A$ and $B$ with $s$ polarization.
    (h) Normalized domain asymmetry
    $\Delta I_{\mathrm{AM}}^{s,\%}$ defined in
    Eq.~\ref{Eq:AM_LP}.
    (i) Same asymmetry with $I_{\mathrm{AM}}^s$ encoded through
    transparency.
    (j--m) Corresponding results for $p$ polarization.
    The dashed circles indicate the angular acceptance
    $\theta=\pm15^\circ$.
    }
         \label{Fig3:LPy} 
    \end{figure}

    The circular-light contrast discussed above reflects both the
    different total photoionization weights and the different angular
    distributions of the $Y_1^{-1}$ and $Y_1^{+1}$ source channels.
    Linear polarization provides a complementary mechanism. For the
    geometry considered here, the integrated atomic photoionization
    intensities of the two channels are identical, whereas their angular
    distributions remain markedly different, as shown in
    Figs.~\ref{Fig3:LPy}~(a--c). Multiple scattering from the inequivalent
    Cr1 and Cr2 environments converts these source-profile differences
    into a domain-dependent PED contrast, without requiring a
    helicity-dependent imbalance of the total emitted intensity.

    Figures~\ref{Fig3:LPy}~(d,e) show the kinetic-energy dependence of the
    atomic photoionization intensities and of the differences between the
    $Y_1^{-1}$ and $Y_1^{+1}$ angular profiles. For circularly polarized
    light, a substantial difference persists over a broad kinetic-energy
    range. For linear polarization, the calculated contrast is strongest
    between approximately $40$ and $100$~eV and decreases gradually at
    higher energies. The proposed mechanism is therefore not restricted to
    a narrowly tuned kinetic energy, although its detailed magnitude will
    depend on the energy-dependent phase shifts and radial matrix elements,
    including possible cross-section anomalies~\cite{Goldberg1981}.
    
    Let us define measurable PED maps from domains $A$ and $B$ for linearly $s$-polarized light, referring to the geometry of Fig. \ref{Fig1:Concept}~(e)
    
    \begin{equation}
    \begin{aligned}
    I^{A,s} &=
    I_{\mathrm{Cr1}}^{A,s}
    +
    I_{\mathrm{Cr2}}^{A,s}, \\
    I^{B,s} &=
    I_{\mathrm{Cr1}}^{B,s}
    +
    I_{\mathrm{Cr2}}^{B,s},
    \end{aligned}
    \end{equation}
    
    Now we isolate the AM contribution as 
    
    \begin{equation}
    \begin{aligned}
    \Delta I_{\mathrm{AM}}^s
    = I^{A,s} - I^{B,s}, \\
    I_{\mathrm{AM}}^s = I^{A,s} + I^{B,s}, \\
    \Delta I_{\mathrm{AM}}^{s,\%} = 100 \cdot \Delta I_{\mathrm{AM}}^s / I_{\mathrm{AM}}^s
    \end{aligned}
         \label{Eq:AM_LP} 
    \end{equation}

    Within the idealized comparison considered here, this domain contrast
    vanishes for a conventional compensated AFM whose opposite magnetic
    sublattices are related by translation and experience equivalent
    surface scattering environments. In CrSb(120), the inequivalent Cr1
    and Cr2 environments allow the reversal of the source-wave character
    between domains $A$ and $B$ to produce a nonzero
    $\Delta I_{\mathrm{AM}}^{s,\%}$.
    
    \section{Discussion}

    The calculated observables connect domain imaging with a test of AM
    order. Their contrast does not arise from conventional dichroism
    alone, but from the correlation between reversal of the
    multiplet-selective source-wave character and the inequivalent Cr1 and
    Cr2 scattering environments. Comparing magnetic domains, light
    polarizations, and the $E_1$ and $E_2$ multiplet regions therefore
    provides mutually constraining signatures that help distinguish
    magnetic contrast from structural PED backgrounds.

    The calculated PED asymmetry exceeds $50\%$ in selected angular
    regions and remains of order $30\%$ within the experimentally relevant
    cone $|\theta|<15^\circ$. The principal experimental uncertainty is
    therefore not the multiple-scattering contrast itself, but the degree
    to which distinct regions of the Cr $3p$ multiplet retain sufficiently
    different $Y_1^m$ source-wave weights. Direct experimental information
    for Cr $3p$ is limited, but magnetic-domain-dependent dichroism
    exceeding $10\%$ was recently observed in the Mn $3p$ multiplet of
    DyMn$_6$Sn$_6$ at $h\nu=300$~eV~\cite{Plucinski2026}. Strong spin
    polarization and magnetic dichroism have also been reported for Fe,
    Co, and Ni $3p$ photoemission
    \cite{Kachel1993,Hillebrecht1995,Rose1996}, while comparatively weak
    linear dichroism was resolved for Ni(110)~\cite{Sacchi1998}. These
    results support the experimental plausibility of resolving
    multiplet-dependent source-wave contrast in transition-metal $3p$
    photoemission.

    The $E_1$ and $E_2$ multiplet regions provide an additional reversal
    operation. For a single domain, however, directly comparing the two
    energies is not by itself conclusive, because their nonmagnetic
    backgrounds and source-channel admixtures need not be identical. Once
    domains $A$ and $B$ have been identified, the difference
    $\Delta I_{\mathrm{AM}}(E_1)-\Delta I_{\mathrm{AM}}(E_2)$ provides a
    stronger internal consistency check. The multiplet-dependent
    source-wave contrast approximately reverses between the two regions,
    whereas backgrounds varying slowly across the Cr $3p$ multiplet,
    including the nearby Sb $4d$ tail and loss satellites, are largely
    suppressed.

    The proposed measurements are compatible with hemispherical analyzers
    equipped with two-dimensional electron deflectors, as well as with
    momentum microscopes available at synchrotron micro-ARPES beamlines.
    Because substantial asymmetry is present within
    $|\theta|<15^\circ$, the essential domain contrast can be accessed
    without recording the full emission hemisphere. The required PED maps
    can therefore be acquired using either momentum microscopes or
    conventional deflector-based analyzers.

    The present calculations describe the Sb-terminated CrSb(120) surface,
    which has been realized experimentally~\cite{Ding2024}. A meaningful
    comparison between domains $A$ and $B$ requires regions with the same
    surface termination, because Sb- and Cr-terminated areas will generally
    produce different nonmagnetic PED backgrounds. In micro-ARPES, the two
    terminations should be distinguishable from the relative Cr $3p$ and
    Sb $4d$ intensities \cite{Li2025}. The most direct experiment would therefore compare
    domains $A$ and $B$ separately on each termination. A spatially stable
    mixture of terminations could in principle also be used, although it
    would introduce an additional source of uncertainty. Thin films may
    require micropatterning if their magnetic domains are smaller than the
    photoemission probe size~\cite{Din2025}. The possible extension to the CrSb(001) and (100) surfaces is briefly considered in the Supplement~\cite{Suppl}.

    The broader significance of the method is that it provides a
    spatially resolved marker of the local N\'eel orientation. Once the
    magnetic domains and surface terminations have been identified,
    valence-band ARPES can be performed on the same microscopic regions
    with a known magnetic reference frame. This enables the
    momentum-dependent spin splitting and orbital character of the
    altermagnetic bands to be correlated with anisotropic transport and
    other spin-dependent electronic responses. Multiplet-selective PED
    therefore connects a core-level diffraction observable to the
    functional electronic properties of altermagnets.
    
    \section{Summary and outlook}

    In summary, we introduced multiplet-selective photoelectron
diffraction as a real-space probe of altermagnetic order and
demonstrated the concept for the CrSb(120) surface. The method exploits
the coupling between multiplet-dependent Cr $3p$ source-wave character
and the inequivalent local scattering environments of the two
antialigned Cr sublattices. Multiple-scattering calculations show that
both circularly and linearly polarized light generate sizable
domain-sensitive PED asymmetries when combined with an appropriate
selection of the Cr $3p$ multiplet region.

The approach differs from conventional core-level magnetic dichroism
because the observable arises from the combined action of
multiplet-selective emission and sublattice-dependent multiple
scattering. It therefore provides element-selective identification of
the local N\'eel orientation and can establish the magnetic reference
frame required to correlate domain-resolved band structure with
anisotropic transport and other spin-dependent electronic responses.

CrSb provides a favorable test case because of its high N\'eel
temperature and experimentally accessible low-index surfaces. More
generally, the approach should apply to altermagnets and compensated
antiferromagnets in which the opposite magnetic sublattices retain
distinguishable local diffraction environments and the relevant
core-level multiplet exhibits magnetic source-wave contrast. MnTe, FeS, and
related transition-metal compounds represent natural candidates for
such studies.
    
    \section{Acknowledgements}
    
    We would like to thank G. Bihlmayer and Y. Mokrousov for fruitful discussions.
    

%

    \clearpage
    \onecolumngrid
    \appendix
    
    \section*{Supplementary Information: Multiplet-Selective Photoelectron Diffraction from an Altermagnet}

    \onecolumngrid
    \setcounter{figure}{0}
    \renewcommand{\thefigure}{S\arabic{figure}}
    \setcounter{equation}{0}
    \renewcommand{\theequation}{S\arabic{equation}}

    \section{Low index CrSb surfaces}

    It is important to take into account that the magnetic order on the surface of an altermagnetic material needs to be analyzed for each surface separately, together with consequences to PED patterns.
    
    The surface labels used throughout this work are defined with respect to the real-space basis vectors $\{a_1,a_2,a_3\}$ shown in Fig.~1~(a) of the main text. The correspondence to conventional Miller--Bravais notation, used e.g. in Ref. \cite{Ding2024}, is:
    \[
    (120) \equiv (01\overline{1}0),
    \qquad
    (100) \equiv (2\overline{1}\overline{1}0).
    \]
    
\begin{figure}[b]
     \centering
         \includegraphics[width=5.5cm]{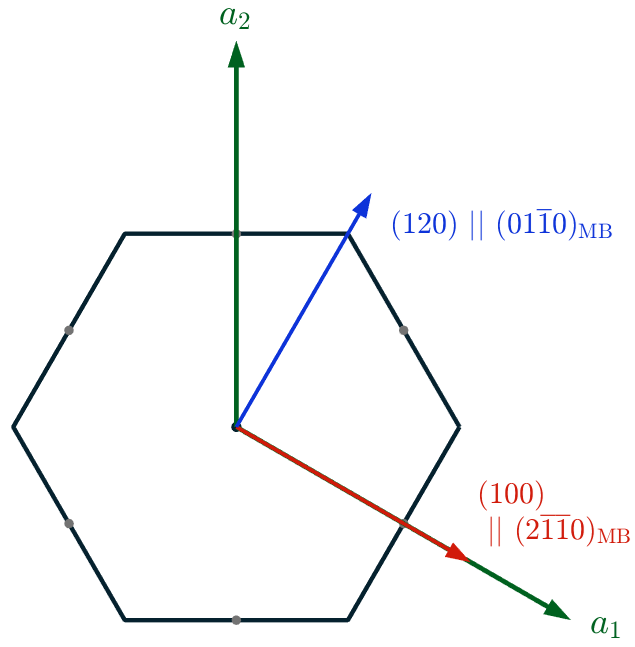}
    \caption{
    Comparison between the surface notation used in this work and the four-index Miller--Bravais notation for hexagonal crystals. Surface normals are expressed with respect to the real-space lattice vectors $a_1$ and $a_2$.
    }
         \label{Supple_Fig:4index} 
    \end{figure}

    The correspondence is illustrated in Fig.~\ref{Supple_Fig:4index}.

The CrSb(001) surface is the most natural cleavage plane and can
exhibit either Cr1 or Cr2 termination, which complicates the study of
AM properties using spin-polarized photoemission. Nevertheless, the
present methodology may be applicable because the local environments
of the two terminations are related by a $60^\circ$ rotation and can
therefore produce different PED responses. The Cr1- and Cr2-terminated
regions must consequently be distinguished experimentally.
    
    If the CrSb(100) side surface can be prepared experimentally \cite{Fischer2026}, the methodology is essentially identical to that described above for CrSb(120).
    
    \begin{figure}[b]
     \centering
         \includegraphics[width=8.5cm]{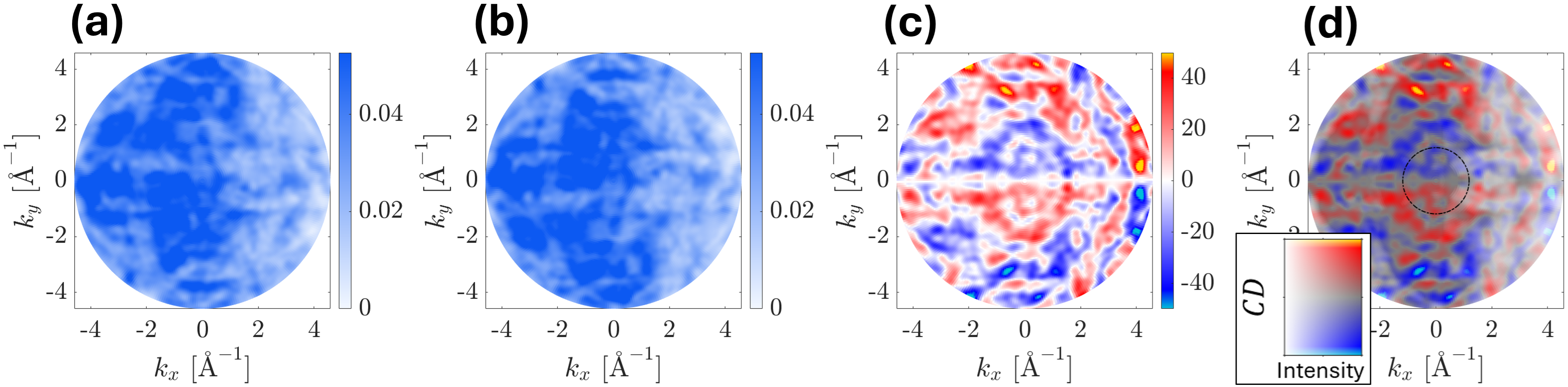}
    \caption{
    Daimon effect at $E_{kin} = 80$ eV. Figure shows the regular CD assuming the perfect $50\%$ mixture of $A$ and $B$ domains. (a) Map with $C_-$ light. (b) Map with $C_+$ light. (c) Circular dichroism map. (d) Map showing both circular dichroism and intensity. See text for details.}
         \label{Supple_Fig:Daimon} 
    \end{figure}
    
    \begin{figure*}
     \centering
         \includegraphics[width=17cm]{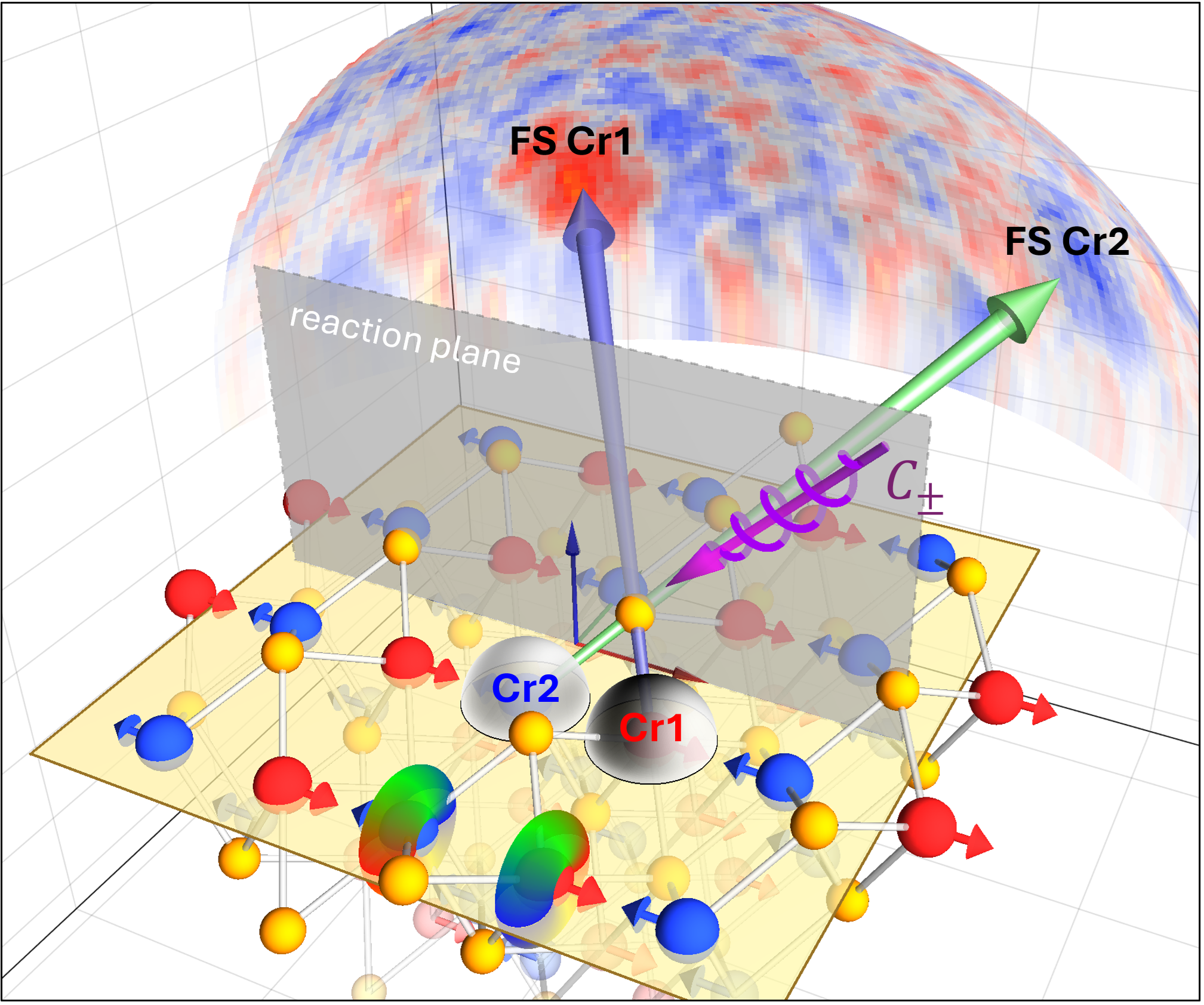}
    \caption{
    Schematic conceptual illustration of probing altermagnetism with PED on the example of CrSb(120) surface. Atomic-like waves emitted from non-equivalent Cr1 and Cr2 sites are forward scattered by the Sb potentials. Two arrows representing this process point to the forward
    scattering peaks FS Cr1 and FS Cr2. These peaks are clearly identified in EDAC \cite{Abajo2001} maps calculated at $E_{kin} = 1$ keV, part of which is also shown as a section of the emission hemisphere.}
         \label{Supple_Fig:FS} 
    \end{figure*}

    \section{Multiplet-selective source-wave picture for Cr \texorpdfstring{$3p$}{3p} photoemission}
    
    The interpretation developed in this work is based on the idea that
    different regions of the Cr $3p$ photoemission spectrum generate
    different effective source waves for the subsequent photoelectron
    diffraction process.
    
    Transition-metal $3p$ photoemission does not produce a simple spin-orbit split doublet. Instead, the final state corresponds to a $3p^5 3d^n$ multiplet manifold shaped by exchange interaction, spin-orbit coupling, and hybridization effects \cite{Kachel1993,Henk1999}. Already early spin-resolved measurements on Fe, Co, and Ni demonstrated that different spectral features of the $3p$ region carry different spin character and exhibit different photoelectron angular distributions~\cite{Kachel1993}.  These effects can be understood qualitatively using simplified methods \cite{Kachel1993,Bethke2005} or numerically \cite{deGroot2005}.

    In the present work, we calculate the Cr $3p$ removal multiplet using
Quanty and analyze its energy-dependent decomposition in the complex
$Y_1^m$ basis. Different energy windows within the multiplet can
preferentially enhance different magnetic quantum numbers $m$, leading
to source waves enriched in $Y_1^{+1}$ or $Y_1^{-1}$ character. The
relevant quantization axis is tied to the local N\'eel vector of the
Cr sublattice.
    
    Within the dipole approximation, the photoemission matrix element can
    be written schematically as
    
    \begin{equation}
    M \propto \langle \psi_f | \mathbf{A}\cdot\mathbf{p} | \phi_i \rangle ,
    \end{equation}
    
    where $\phi_i$ denotes the localized Cr $3p$ initial state and
    $\psi_f$ the outgoing photoelectron state.
    For circularly polarized light, the operator
    $\mathbf{A}\cdot\mathbf{p}$ naturally couples to combinations of spherical harmonics with opposite angular momentum projections, favoring different $m$ channels for opposite helicities. However, because the Cr $3p$ multiplet already contains strongly mixed
    spin-orbital character, changing the energy window within the multiplet can produce effects analogous to reversing the helicity itself.
    
    The emitted photoelectron subsequently undergoes multiple
    scattering in the surrounding crystal environment.
    This step is crucial because the experimentally observed angular
    patterns are dominated not only by the initial atomic photoionization
    process but also by photoelectron diffraction and interference effects. In particular, circular dichroism in angular distributions can arise
    even in nonmagnetic systems through the Daimon
    effect~\cite{Daimon1993}.

Therefore, observing a dichroic PED pattern alone is not sufficient
evidence for altermagnetic order. This is illustrated in
Fig.~\ref{Supple_Fig:Daimon}, where we calculate the circular
dichroism after summing the contributions from domains $A$ and $B$:

\begin{equation}
\begin{aligned}
\Delta I_{\mathrm{CD}}
&= I^{A,C_-} + I^{B,C_-}
 - I^{A,C_+} - I^{B,C_+}, \\
I_{\mathrm{SUM}}
&= I^{A,C_-} + I^{B,C_-}
 + I^{A,C_+} + I^{B,C_+}, \\
\Delta I_{\mathrm{CD}}^\%
&= 100 \cdot \Delta I_{\mathrm{CD}} / I_{\mathrm{SUM}} .
\end{aligned}
\label{Eq:CD}
\end{equation}

The resulting patterns exhibit strong nonmagnetic dichroic contrast.

    The central idea of the present work is that the magnetic contribution
can be isolated through correlated reversal operations. If a component
of the dichroic diffraction signal reverses consistently under either
N\'eel-domain reversal or multiplet-energy-window reversal, this
supports its association with altermagnetic order. Figure
\ref{Supple_Fig:FS} provides an intuitive illustration of the processes
involved and extends Fig.~1(e) of the main text. It depicts forward
scattering of waves emitted from the Cr1 and Cr2 sites by the surface
Sb potential. At high kinetic energies, this process produces
forward-scattering features with strong circular dichroism.
    
    This framework is conceptually connected to earlier studies of magnetic dichroism and spin-polarized photoelectron diffraction
    \cite{Sinkovic1985,Fadley1997}, but here extended toward the case of
    compensated altermagnetic order and multiplet-selective real-space
    contrast at a single magnetic domain.

    \section{Additional PED calculations}

\begin{figure}
     \centering
         \includegraphics[width=8.5cm]{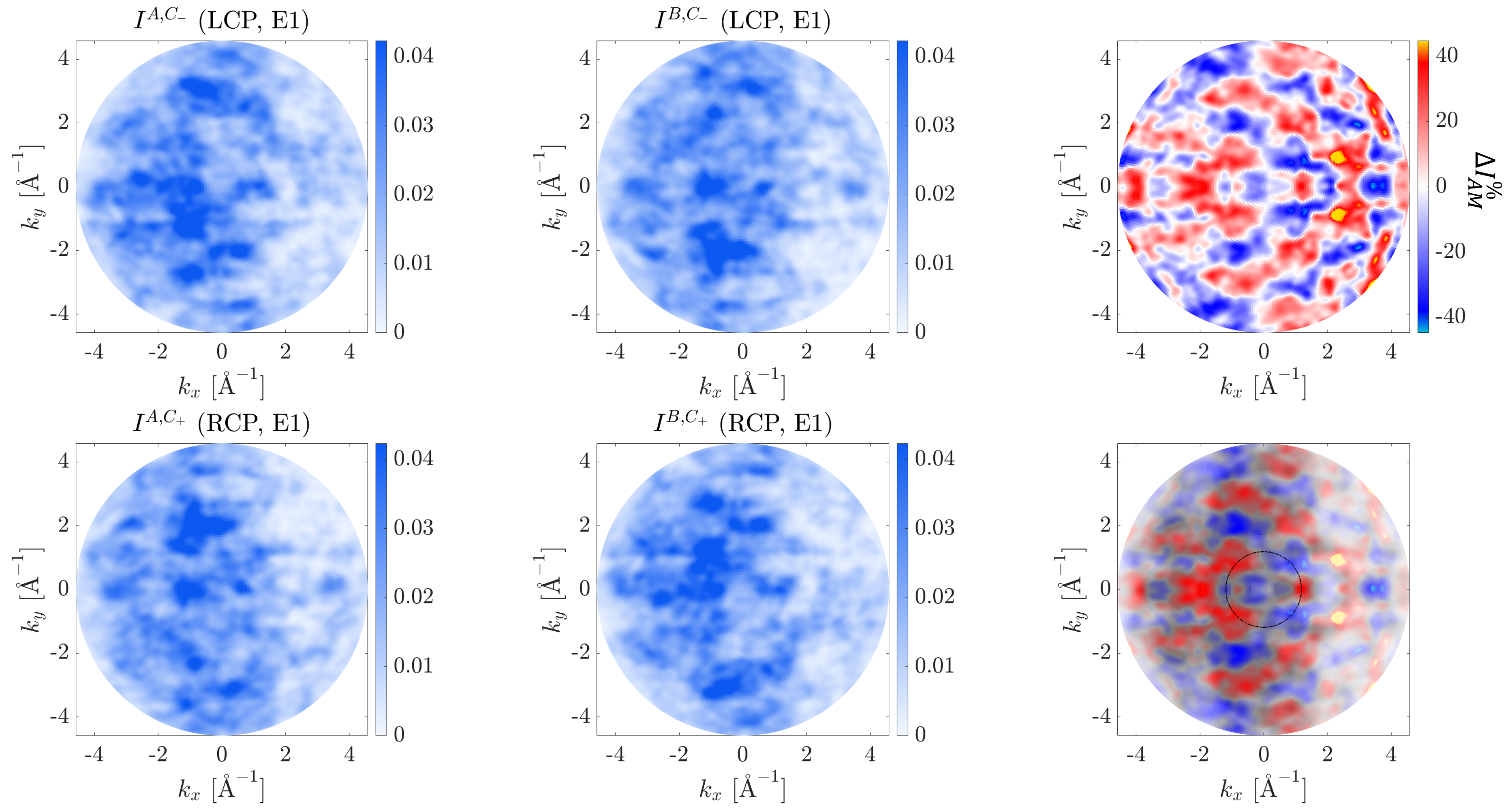}
    \caption{
    Same as Fig. 2 of the main text, but summing emissions from three outermost Cr layers, $E_{kin} = 80$ eV.}
         \label{Supple_Fig:three_layers} 
    \end{figure}

In the main text, only emission from the outermost Cr layer was
included in the PED calculations. Figure
\ref{Supple_Fig:three_layers} shows the corresponding calculation with
the three outermost Cr layers included, comprising six emitting sites
in total. The inclusion of the deeper layers does not significantly
change the magnitude of the calculated effect.
    
\begin{figure}
     \centering
         \includegraphics[width=8.5cm]{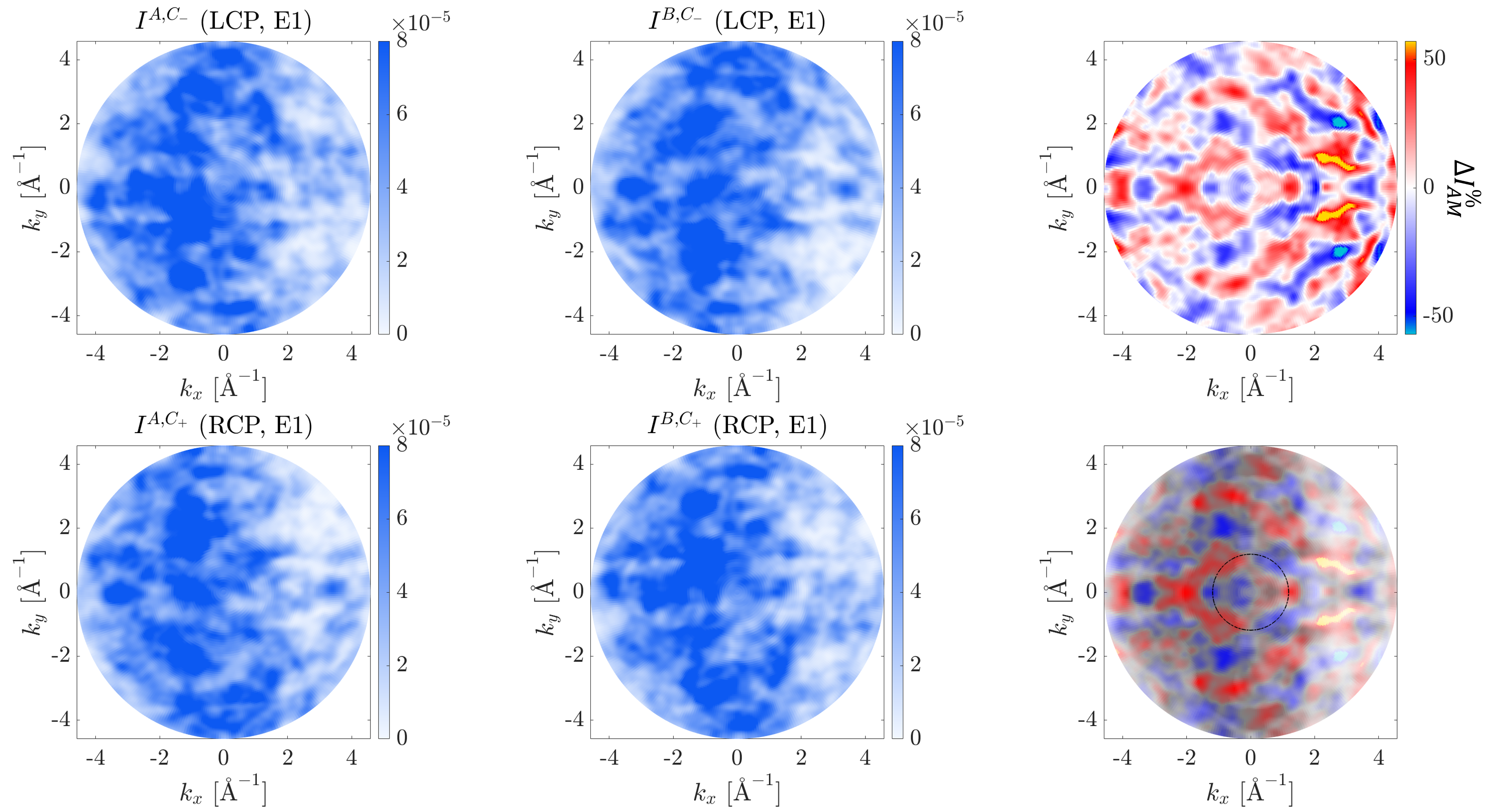}
    \caption{
    Same as Fig. 2 of the main text, $E_{kin} = 80$ eV, but using the Clementi-Roetti \cite{Clementi1974} initial wave function.}
         \label{Supple_Fig:C-R} 
    \end{figure}

In the photoemission calculations performed with EDAC
\cite{Abajo2001}, different initial-state radial wave functions can be
used. In the main text, we use the default wave function calculated
internally by EDAC. As an alternative, we employ the
Roothaan--Hartree--Fock atomic wave functions tabulated by
Clementi and Roetti~\cite{Clementi1974}. Figure
\ref{Supple_Fig:C-R} shows PED patterns calculated using the
Clementi--Roetti Cr $3p$ wave function for a neutral Cr atom in the
$^7S$ atomic-term configuration. The result is nearly identical to
Fig.~2 of the main text, indicating that the precise radial form of
the initial-state wave function does not significantly affect the PED
maps.

The radial wave function does, however, affect the kinetic-energy
dependence of the atomic photoionization profiles
\cite{YehLindau}, including the possible occurrence of Cooper minima.
The Clementi--Roetti radial function was therefore used for
Figs.~3(d,e) of the main text.

\begin{figure}
     \centering
         \includegraphics[width=8.5cm]{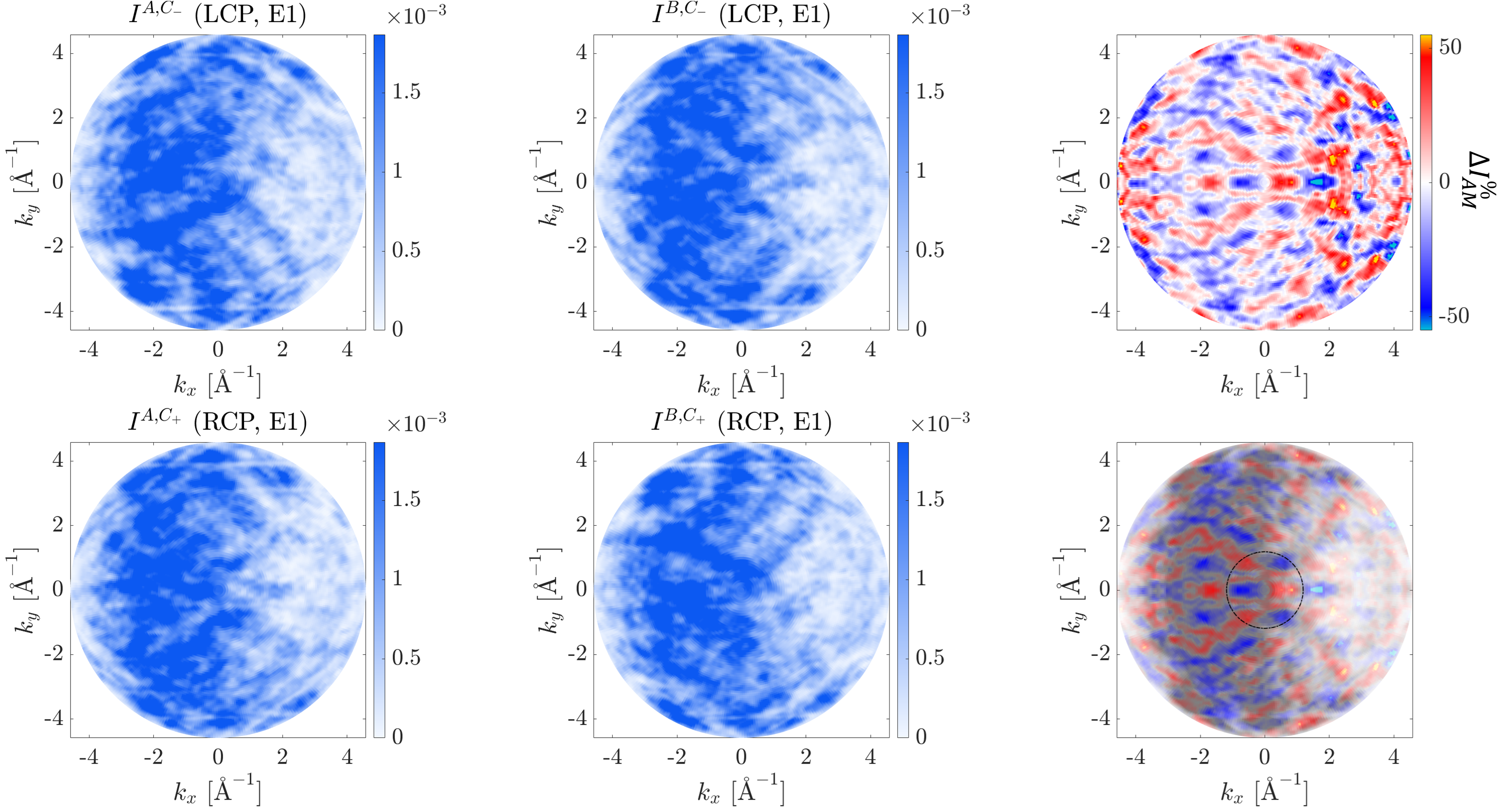}
    \caption{
    Same as Fig. 2 of the main text but for $E_{kin} = 250$ eV.}
         \label{Supple_Fig:250} 
    \end{figure}

\begin{figure}
     \centering
         \includegraphics[width=8.5cm]{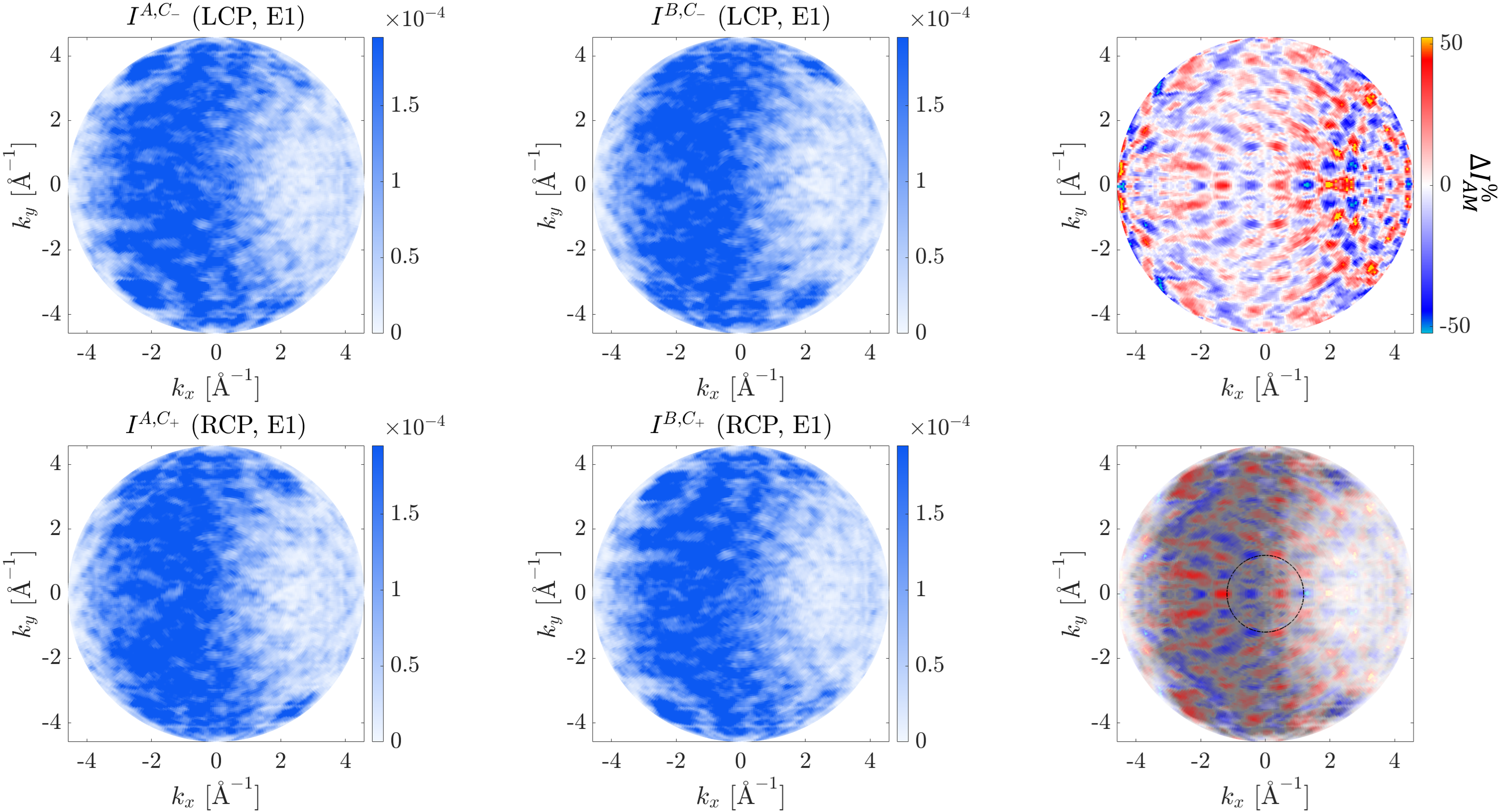}
    \caption{
    Same as Fig. 2 of the main text but for $E_{kin} = 500$ eV.}
         \label{Supple_Fig:500} 
    \end{figure}

\begin{figure}
     \centering
         \includegraphics[width=8.5cm]{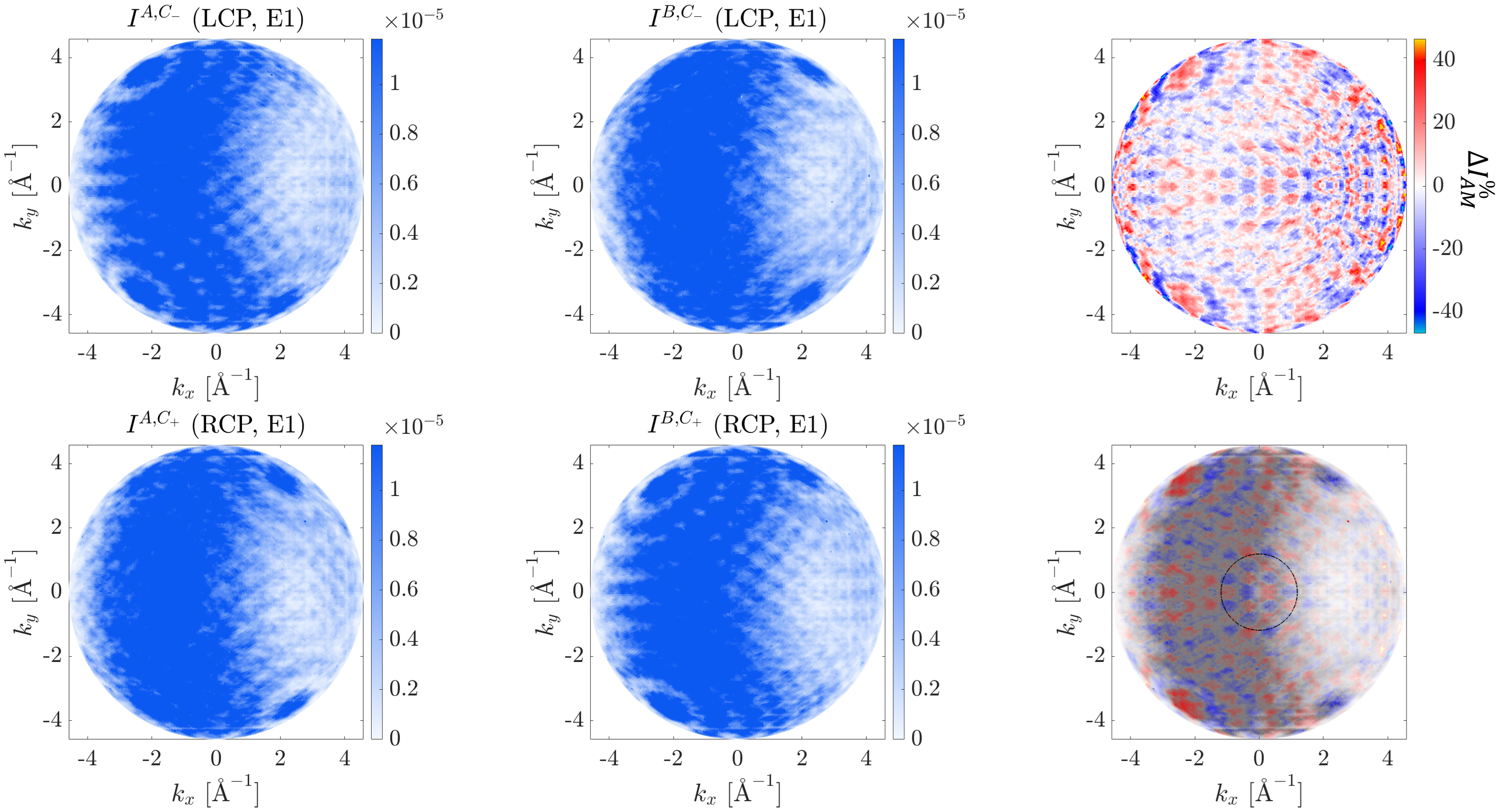}
    \caption{
    Same as Fig. 2 of the main text but for $E_{kin} = 1000$ eV.}
         \label{Supple_Fig:1000} 
    \end{figure}

    Figures \ref{Supple_Fig:250}, \ref{Supple_Fig:500}, and \ref{Supple_Fig:1000} show additional EDAC results at $E_{kin}$ of 250 eV, 500 eV, and 1000 eV. The forward scattering features depicted in Fig. \ref{Supple_Fig:FS} can be clearly seen at the $1000$ eV maps.

    \section{Proposed experimental procedures to determine AM order}

    In practice, measurements with a hemispherical analyzer equipped with
angular deflectors could be performed at a micro-ARPES beamline with
sufficient photon-beam focusing. First, the surface Brillouin-zone
orientation should be determined from ARPES deflector scans acquired
on a region exhibiting sharp spectral features and low background.
The sample should then be aligned, for example, according to the
geometry shown in Fig.~\ref{Supple_Fig:FS}. The analyzer energy window
should include the Cr $3p$ and Sb $4d$ regions, while lateral maps are
acquired by rastering the sample in fixed mode. This mode allows the
entire Cr $3p$ region to be recorded simultaneously. Modern
hemispherical analyzers can acquire an $E(k_{\parallel})$ map over an
angular range of approximately $\pm15^\circ$ in a single exposure.
The analyzer slit direction should be oriented perpendicular to the
reaction plane. A complete two-dimensional deflector map can also be
measured, although this would be considerably more time-consuming
during sample rastering.

This procedure yields data sets
$I(E,x,y,k_{\parallel})$ for different light polarizations, where
$x$ and $y$ denote the two lateral surface coordinates. The relative
Cr $3p$ and Sb $4d$ intensities can be used to distinguish regions
with different surface terminations. After regions with the same
termination have been identified, the quantities
$\Delta I_{\mathrm{AM}}^\%$ and
$\Delta I_{\mathrm{AM}}^{s,\%}$ defined in the main text can be
constructed to locate the AM domains. Spin-resolved ARPES can then be
performed on individual domains to determine the spin-polarized
surface and bulk-projected band structures.

  A similar procedure can be applied to measurements performed with a
momentum microscope. In this case, it is advantageous to identify the
$E_1$ and $E_2$ regions of the Cr $3p$ multiplet in advance, where the
polarization-dependent contrast is expected to be strongest, because
scanning the photoelectron kinetic energy is comparatively
time-consuming with momentum microscopes.

    \section{Remarks on PED theory}

Within the dipole approximation, the photoemission intensity is
proportional to $|M|^2$, where

\begin{equation}
M(\mathbf{k}_f)
\propto
\left\langle
\psi_f(\mathbf{k}_f)
\left|
\mathbf{A}\cdot\mathbf{p}
\right|
\psi_i
\right\rangle .
\end{equation}

Here $\mathbf{k}_f$ specifies the photoelectron kinetic energy and
emission direction. The initial state may be expanded in a localized
orbital basis as

\begin{equation}
\psi_i=\sum_j C_j\phi_j ,
\end{equation}

where $j$ labels the atomic site and orbital quantum numbers. For
valence-band states, the coefficients $C_j$ generally depend on the
initial-state wave vector and the corresponding amplitudes must be
summed coherently. For localized core-level photoemission, emissions
associated with distinct core-hole sites are commonly treated as
incoherent, while the angular-momentum channels associated with a
given emitter remain coherent. The measured intensity can therefore
be written schematically as

\begin{equation}
I(\mathbf{k}_f)
\propto
\sum_{\alpha}
\left|
\left\langle
\psi_f(\mathbf{k}_f)
\left|
\mathbf{A}\cdot\mathbf{p}
\right|
\psi_{i,\alpha}
\right\rangle
\right|^2 ,
\end{equation}

where $\alpha$ labels the distinct localized emitting sites.

    In PED one often uses a picture of a spherical wave being emitted from one site and being scattered by neighboring sites \cite{Fadley1997}, which somewhat resembles time-dependent DFT attempts for calculating ARPES spectra \cite{DeGiovannini2016}.

In PED, the photoemission final state is commonly treated as a
time-reversed LEED state~\cite{Breit1954}. The intuitive picture of a
wave emitted from one site and subsequently scattered by neighboring
atoms~\cite{Fadley1997} provides a useful representation of the
corresponding multiple-scattering process.

The PED calculations in this work use the multiple-scattering
formalism implemented in EDAC~\cite{Abajo2001}. Its principal
approximations include muffin-tin scattering potentials and a
step-like surface barrier, rather than a fully self-consistent surface
potential~\cite{Henk1993,Braun1996}.
    
    Nevertheless, the intuitive emitted-and-scattered-wave picture remains extremely useful for understanding the origin of PED contrast and the emergence of forward-scattering features in momentum-space maps, provided that the underlying calculations are performed within the full multiple-scattering formalism.

    \begin{figure}
     \centering
     \includegraphics[width=8.5cm]{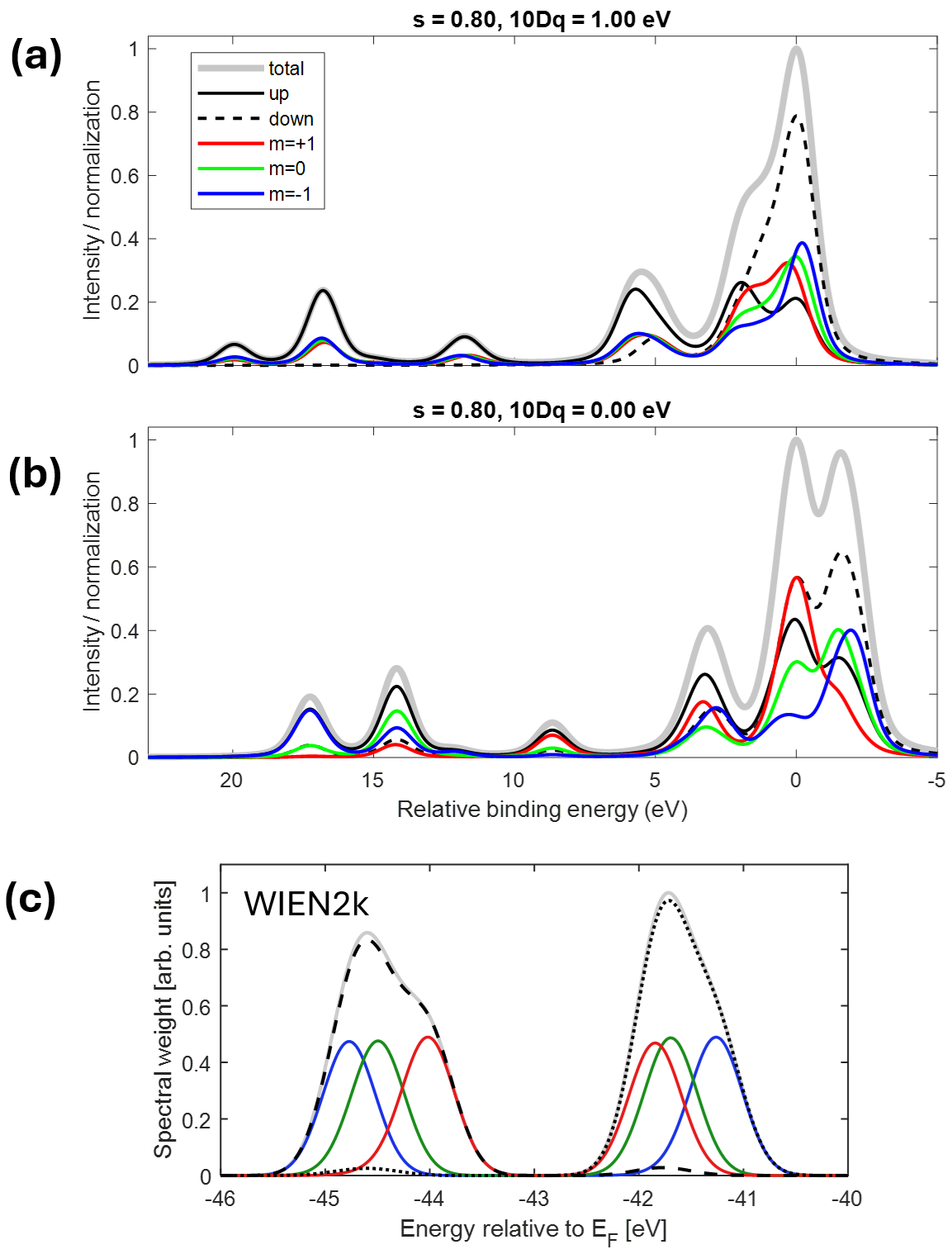}
    \caption{
    Comparison of many-body and one-electron descriptions of the Cr $3p$
    region.
    (a,b) Atomic Cr$^{3+}$ multiplet spectral functions calculated using
    Quanty~\cite{Haverkort2014} for the
    $3p^6 3d^3\rightarrow3p^5 3d^3$ photoemission process. The Slater
    integrals were obtained from FPLO atomic radial functions
    \cite{Koepernik1999,FPLO2202} and reduced uniformly to $80\%$ of
    their bare atomic values \cite{deGroot2001,Wray2012}.
    Panel (a) shows the reference calculation with an octahedral
    crystal-field splitting $10Dq=1.0$~eV, while panel (b) shows the
    spherical limit with $10Dq=0$. In each panel, the thick gray curve
    denotes the total spectral function, the solid and dashed black curves
    denote the spin-up and spin-down contributions, and the red, green,
    and blue curves denote the spin-summed $Y_1^{+1}$, $Y_1^0$, and
    $Y_1^{-1}$ components, respectively. The spectra are normalized
    individually and their energy axes are referenced to the maximum of
    the principal multiplet feature.
    (c) One-electron WIEN2k Cr $3p$ spectral weights
    \cite{Blaha2020}, decomposed into spin
    and complex $Y_1^m$ components. For visualization, the discrete WIEN2k levels were broadened with a Gaussian of $\sigma=0.25$~eV, corresponding to a full width at half maximum of
    approximately $0.59$~eV. The one-electron calculation produces
discrete exchange- and spin--orbit-split levels but does not reproduce
the many-body multiplet envelope, it is shown only as an illustrative
single-particle comparison.
    }
    \label{Supple_Fig:Quanty}
    \end{figure}

    \section{Computational parameters}
    
    \subsection{FPLO}
    
    The Cr $3p$ multiplet and its energy-dependent source-wave
    decomposition were calculated using an atomic many-body model.
    Atomic radial functions and interaction parameters were obtained using
    the full-potential local-orbital (FPLO) code
    \cite{Koepernik1999}, version 22.02-66
    \cite{FPLO2202}. Spherically averaged atomic Cr$^{3+}$
    calculations were performed for the $3d^3$ configuration. The
    nonrelativistic radial functions were used to evaluate the Slater
    integrals, while a fully relativistic atomic calculation was used to
    determine the Cr $3p$ and $3d$ spin--orbit coupling parameters.
    
    The resulting bare atomic interaction parameters were
    
    \begin{equation}
    \begin{aligned}
    F^2_{dd} &= 10.665~\mathrm{eV}, &
    F^4_{dd} &= 6.664~\mathrm{eV},\\
    F^2_{pd} &= 11.560~\mathrm{eV}, &
    G^1_{pd} &= 14.343~\mathrm{eV},\\
    G^3_{pd} &= 8.718~\mathrm{eV}.
    \end{aligned}
    \end{equation}
    
    The fully relativistic atomic calculation yielded
    
    \begin{equation}
    \zeta_{3p}=0.720~\mathrm{eV},
    \qquad
    \zeta_{3d}=0.0445~\mathrm{eV}.
    \end{equation}
    
    To account phenomenologically for solid-state screening and
    configuration-interaction effects not included explicitly in the
    restricted atomic model, all multipolar Slater integrals were reduced
    uniformly to $80\%$ of their atomic values, consistent with the
    conventional scaling used in transition-metal multiplet calculations
    \cite{deGroot2001,Wray2012}. The spin--orbit coupling parameters were
    left unscaled. The parameters used in the reference calculation were
    therefore
    
    \begin{equation}
    \begin{aligned}
    F^2_{dd} &= 8.532~\mathrm{eV}, &
    F^4_{dd} &= 5.331~\mathrm{eV},\\
    F^2_{pd} &= 9.248~\mathrm{eV}, &
    G^1_{pd} &= 11.474~\mathrm{eV},\\
    G^3_{pd} &= 6.974~\mathrm{eV}.
    \end{aligned}
    \end{equation}
    
    \subsection{Quanty}
    
    The Cr $3p$ spectral function was calculated using
    Quanty~\cite{Haverkort2014} for the photoemission process
    
    \begin{equation}
    3p^6 3d^3
    \longrightarrow
    3p^5 3d^3 .
    \end{equation}
    
    In the reference calculation, a minimal octahedral crystal field with
    $10Dq=1.0$~eV was introduced to produce an orbitally quenched,
    high-spin $t_{2g}^3$ initial state. The calculated state had
    
    \begin{equation}
    \langle S_d^2\rangle = 3.748,
    \end{equation}
    
    close to the ideal $S=3/2$ value
    $S(S+1)=3.75$, and a $t_{2g}$ occupation of approximately $2.996$.
    
    The octahedral field is used here as a minimal phenomenological
    representation of orbital quenching and is not intended as a
    material-specific determination of the crystal-field parameters in
    CrSb. As shown in Fig.~\ref{Supple_Fig:Quanty}, the calculation with
    $10Dq=1.0$~eV produces a less structured principal
    multiplet envelope than the spherical $10Dq=0$ limit, qualitatively
    closer to the weakly structured Cr $3p$ photoemission spectrum (see e.g. the supplement in Ref. \cite{Li2025}).
    Nevertheless, the spherical calculation was retained as an explicit
    robustness test.
    
    The inclusion of a crystal field is also conceptually connected to
    linear dichroism in transition-metal core-level spectroscopy. A
    non-spherical crystal field produces an anisotropic local electronic
    structure that can be probed using differently oriented linear
    polarizations, giving rise to crystal-field linear dichroism
    \cite{StohrSiegmann2006,Haverkort2004}. This contribution is distinct
    from magnetic linear dichroism, although the two can coexist and may
    produce similar spectral signatures. Their separation generally
    requires an explicit treatment of the polarization-dependent dipole
    matrix elements and experimental geometry
    \cite{Haverkort2004}.
    
    In the present calculation, the crystal field modifies the intrinsic
    Cr $3p$ removal multiplet and its $Y_1^m$ composition. The curves in
    Fig.~\ref{Supple_Fig:Quanty}~(a,b), however, are projections of the
    polarization-independent spectral density matrix and do not themselves
    represent a calculated linear-dichroism spectrum. 
    
    The monopole parameters $F^0_{dd}$ and $F^0_{pd}$ were set to zero.
    Within the fixed initial and final charge sectors considered here,
    these terms produce only constant energy shifts and do not affect the
    relative multiplet splittings or the source-wave composition. An
    infinitesimal spin-selecting field of $10^{-5}$~eV was used only to
    select one member of the initial magnetic multiplet and was not
    included in the physical final-state Hamiltonian.
    
    The calculated spectral functions used a Lorentzian broadening
    $\Gamma=0.5$~eV and were subsequently convoluted with a Gaussian of
    $1.0$~eV full width at half maximum.
    
    Quanty \cite{Haverkort2014} was used to calculate the full $6\times6$ Green-function tensor
    in the Cr $3p$ removal basis $|Y_1^m,\sigma\rangle$. The corresponding
    energy-resolved spectral density matrix was evaluated as
    
    \begin{equation}
    \rho(E)
    =
    -\frac{G(E)-G^\dagger(E)}{2\pi i}.
    \end{equation}
    
    The spin-summed orbital contributions were obtained from the diagonal
    elements,
    
    \begin{equation}
    w_m(E)
    =
    \sum_{\sigma}
    \rho_{m\sigma,m\sigma}(E),
    \qquad
    m=-1,0,+1.
    \end{equation}
    
    These quantities characterize the intrinsic $Y_1^m$ composition of the
    Cr $3p$ removal spectrum. They should not be identified directly with
    the measured photoemission intensity, which additionally depends on
    the light polarization, emission direction, dipole matrix elements,
    and final-state multiple scattering.
    
    Reversal of the local magnetic domain acts as time reversal on the
    local source state,
    
    \begin{equation}
    (m,\sigma)\longrightarrow(-m,-\sigma).
    \end{equation}
    
    After tracing over spin, this gives
    
    \begin{equation}
    w_m^{B}(E)=w_{-m}^{A}(E),
    \end{equation}
    
    so that the $Y_1^{+1}$ and $Y_1^{-1}$ source-wave contributions are
    interchanged between the two domains, whereas the $Y_1^0$
    contribution remains unchanged.
    
    The atomic calculation is not intended to reproduce quantitatively the
    complete experimental Cr $3p$ line shape. In metallic CrSb,
    hybridization, charge transfer, nonlocal screening, and
    configuration-dependent relaxation are not fully represented by the
    restricted
    $3p^6 3d^3\rightarrow3p^5 3d^3$
    model. Its purpose is instead to identify energy regions within the
    principal multiplet envelope that are enriched predominantly in
    $Y_1^{-1}$ or $Y_1^{+1}$ source-wave character.
    
    The robustness of this decomposition was tested by varying the common
    Slater-integral reduction factor between $0.70$ and $0.90$ and the
    crystal-field splitting between $10Dq=0$ and $1.5$~eV. These
    variations modify the detailed multiplet structure, particularly the
    high-binding-energy satellites. The principal multiplet envelope and
    the qualitative interchange of the $Y_1^{+1}$ and $Y_1^{-1}$
    contributions, however, remain present. The calculations with
    $10Dq=1.0$~eV and $10Dq=0$ are compared explicitly in
    Figs.~\ref{Supple_Fig:Quanty}~(a) and
    \ref{Supple_Fig:Quanty}~(b), respectively.
    
    For comparison, a one-electron Cr $3p$ core-level decomposition was
    also obtained from a spin-polarized density-functional calculation
    using WIEN2k~\cite{Blaha2020}. The calculation produces discrete
    exchange- and spin--orbit-split one-electron levels but cannot describe
    the many-body $3p^5 3d^3$ final-state manifold. The WIEN2k result,
    shown in Fig.~\ref{Supple_Fig:Quanty}~(c), is included only to
    illustrate the limitations of the single-particle picture. It was not
    used to define the $E_1$ and $E_2$ regions or as input to the PED
    calculations.
    
    For completeness, the one-electron WIEN2k result can be reproduced
    qualitatively by a minimal $LS$-basis model for the Cr $3p$ shell,
    
    \begin{equation}
    H_{\mathrm{toy}}
    =
    \lambda\,\mathbf{L}\!\cdot\!\mathbf{S}
    +
    \Delta_{\mathrm{ex}} S_z ,
    \end{equation}
    
    written in the basis $|Y_1^m,\sigma\rangle$, with
    $m=-1,0,+1$ and $\sigma=\uparrow,\downarrow$. Using
    
    \begin{equation}
    \lambda=0.70~\mathrm{eV},
    \qquad
    \Delta_{\mathrm{ex}}=2.70~\mathrm{eV},
    \end{equation}
    
    the model reproduces the qualitative ordering and approximate energy
    separations of the exchange- and spin--orbit-split Cr $3p$ levels
    obtained from WIEN2k. Here $\lambda$ describes the $3p$ spin--orbit
    interaction, while $\Delta_{\mathrm{ex}}$ represents an effective
    local exchange field. In the absence of spin--orbit coupling, the
    latter separates the pure spin-up and spin-down states by
    approximately $\Delta_{\mathrm{ex}}$. This toy model is included only
    as an instructive one-electron reference.

    As an independent check of the FPLO-to-Quanty workflow, we applied
the same procedure to the well-studied Mn$^{2+}$
$3p^6 3d^5\rightarrow3p^5 3d^5$ problem. In that case, the FPLO spin--orbit
parameters, $\zeta_{3p}=0.852$~eV and $\zeta_{3d}=0.0497$~eV, are
close to the values $0.79$ and $0.046$~eV used in recent Mn $3p$
photoemission modeling~\cite{Plucinski2026}. With the same $80\%$
reduction of the FPLO Slater integrals, the Quanty calculation produces
the characteristic principal multiplet together with a remote
high-binding-energy recoupling branch, consistent with the established
atomic interpretation of Mn $3p$ photoemission
\cite{Bagus2000}. This benchmark supports the physical scale of the
extracted atomic parameters and the internal consistency of the
workflow.

    \subsection{EDAC}
    
    PED calculations were carried out using the EDAC multiple-scattering
    code~\cite{Abajo2001} with the spherical cluster of 1693 atoms shown
    in Fig.~1(g) of the main text. Unless stated otherwise, the
    calculations employed $l_{\mathrm{max}}=20$, scattering order 20, an
    inner potential $V_0=15$~eV, and an inelastic mean free path of
    5~\AA. The latter value is physically reasonable near
    $E_{\mathrm{kin}}\approx80$~eV but becomes unrealistically small at
    substantially higher kinetic energies. Consequently, the
    higher-energy PED simulations shown in
    Figs.~\ref{Supple_Fig:500} and \ref{Supple_Fig:1000} should be
    regarded primarily as qualitative illustrations of forward-scattering
    behavior.

    \end{document}